\documentclass[runningheads]{llncs}

\usepackage{orcidlink}
\usepackage{listings}
\usepackage{caption}
\usepackage{courier}
\usepackage{tabularx}
\usepackage{multirow}
\usepackage{booktabs}
\usepackage{makecell}
\usepackage{array}
\usepackage{ragged2e}
\usepackage{makecell}
\usepackage[table]{xcolor}
\usepackage{tabularx}
\usepackage{cite}
\usepackage{amsmath,amssymb,amsfonts}
\usepackage{pifont}
\usepackage{algorithmic}
\usepackage{graphicx}
\usepackage{textcomp}
\usepackage{xcolor}
\makeatletter
\def\input@path{{./}}
\makeatother
\usepackage[most]{tcolorbox}
\usetikzlibrary{calc}
\usepackage{tikz}
\usetikzlibrary{arrows.meta, positioning}
\usepackage{listings}

\lstdefinestyle{jsonstyle}{
	basicstyle=\ttfamily\small,
	breaklines=true,
	breakatwhitespace=true,
	showstringspaces=false,
	frame=none,
	columns=fullflexible,
	keepspaces=true,
	escapeinside={\%\%}{\%\%},
	stringstyle=\color{brown},
	keywordstyle=\color{blue},
	commentstyle=\color{gray}
}

\def\BibTeX{{\rm B\kern-.05em{\sc i\kern-.025em b}\kern-.08em
    T\kern-.1667em\lower.7ex\hbox{E}\kern-.125emX}}

\begin{document}

\title{WiFiPenTester: Advancing Wireless Ethical Hacking with Governed GenAI}

\titlerunning{Introducing WiFiPenTester}

\author{
	Haitham S.\ Al-Sinani\inst{1,3}\orcidlink{0009-0005-0453-3335}
	\and Chris J.\ Mitchell\inst{2}\orcidlink{0000-0002-6118-0055}
}

\institute{
	Diwan of Royal Court, Muscat,  Oman.  \email{hsssinani@diwan.gov.om} 
	\and Royal Holloway, University of London, Egham, UK. 	\email{C.Mitchell@rhul.ac.uk} 
	\and German University of Technology in Oman, Muscat,  Oman.  \email{Haitham.AlSinani@gutech.edu.om}
}

\authorrunning{H. Al-Sinani \& C. Mitchell}

\maketitle
\begin{abstract}
	Wireless ethical hacking  relies heavily on skilled practitioners manually interpreting reconnaissance results and executing complex, time-sensitive sequences of commands to identify vulnerable targets, capture authentication handshakes, and assess password resilience; a process that is inherently labour-intensive, difficult to scale, and prone to subjective judgement and human error. To help address these limitations, we propose  \texttt{WiFiPenTester}, an experimental, governed, and reproducible system for GenAI-enabled wireless ethical hacking. The system integrates large language models  into the reconnaissance and decision-support phases of wireless security assessment, enabling intelligent target ranking, attack feasibility estimation, and strategy recommendation, while preserving strict human-in-the-loop  control and budget-aware execution. We describe the system architecture, threat model, governance mechanisms, and prompt-engineering methodology, and   empirical experiments  conducted across multiple  wireless environments. The results demonstrate that GenAI assistance improves target selection accuracy and overall assessment efficiency, while maintaining auditability and ethical safeguards. This indicates that \texttt{WiFiPenTester} is a meaningful step toward practical, safe, and scalable GenAI-assisted wireless penetration testing, while reinforcing the necessity of bounded autonomy, human oversight, and rigorous governance mechanisms when deploying GenAI in ethical hacking.
	\keywords{AI \and GenAI \and Wireless Security \and Ethical Hacking  \and LLM \and Human-in-the-Loop \and Penetration Testing}
\end{abstract}

\section{Introduction}
\label{Introduction}

Wireless penetration testing (PenTesting)  is essential for assessing the security posture of modern IEEE~802.11 networks, underpinning enterprise connectivity, public infrastructure, and critical services. Despite significant improvements since the era of WEP and early WPA,  wireless deployments continue to suffer from weak or reused pre-shared keys, insecure client behaviour, poor access point (AP) configuration, residual support for legacy protocols, and insufficient monitoring of association and deauthentication events. Wireless PenTesting, therefore, remains a critical component of organisational security assurance.

In practice, wireless ethical hacking is heavily dependent on skilled practitioners manually interpreting reconnaissance results, selecting viable targets, configuring monitor-mode interfaces, orchestrating active deauthentication and handshake capture procedures, and assessing password resilience through offline cracking. These tasks must be performed in dynamic radio-frequency (RF) environments where client presence, channel conditions, and signal quality fluctuate continuously, requiring substantial expertise and situational awareness. As a result, wireless security assessment is labour-intensive, difficult to scale, and prone to subjective judgement and operational error.

To reduce operator burden and improve consistency, existing toolchains, such as \texttt{Aircrack-ng}\footnote{\url{https://www.aircrack-ng.org/}}~\cite{2022_AircrackNG_ChristopheDevine} and \texttt{Wifite}\footnote{\url{https://www.kali.org/tools/wifite/}}~\cite{2020_wifite2_Deric}, provide scripted workflows for scanning, handshake capture, and dictionary-based key recovery. While effective, such tools rely almost entirely on static heuristics or human intuition for target selection, attack feasibility estimation, and strategy configuration, and offer limited support for structured reasoning, auditability, or reproducibility of decision processes. As wireless environments become denser and more heterogeneous, these limitations increasingly constrain both research experimentation and professional PenTesting practice.

Recent advances in Generative Artificial Intelligence (GenAI), particularly Large Language Models (LLMs), have demonstrated strong capabilities in contextual reasoning, structured decision support, and adaptive strategy generation across a wide range of domains. In cybersecurity, early studies have explored LLMs for vulnerability analysis, malware classification, exploit generation, and automated PenTesting workflows~\cite{
	PSTPaper_PenTestModel_2010,
	YaroslavStefinko_ManualAndAutomatedPenetrationTesting_2016,
	2018_AutomatedPenetrationTestingAnOverview_FarahAbuDabaseh,
	2023_GettingGettingPwndbyAI_PenTestingWithLLMs_AndreasHappe,
	2024_ComprehensiveOverviewLLMs_MohammedHassanin,
	2024_PentestGPT_Deng,
	2025_Penterep_WilliLazarov,
	STM24_UnleashingAIinEthicalHacking,
	TechReportUnAIInEH_HC_2024,
	ITASEC25Paper_AdvancingEthicalHackingWthAIALinux-basedExperimentalStudy_2025,
	2024_APracticalExaminationofManualExploitationPrivilegeEscalation_HC,
	2025_PenTest++_HC_CyBAI,
	PenTest++_2025pentestelevatingethicalhacking_HaithamChris,
	PenTest20_2025autonomousprivesc_HaithamChris,
	2026_PenTest20_AINA26_HaithamChris
}. However, the integration of GenAI into live wireless PenTesting remains largely under-explored, particularly in settings that demand strict safety guarantees, legal compliance, cost control, and human oversight.

To address this gap, we present \texttt{WiFiPenTester}, a  system for governed and reproducible GenAI-enabled wireless ethical hacking. The system integrates LLM-based reasoning into the reconnaissance and decision-support phases of wireless security assessment, enabling intelligent target ranking, attack feasibility estimation, and strategy recommendation from structured scan metadata, while preserving strict human-in-the-loop (HITL) control over all active and potentially disruptive operations. Unlike prior automation tools, \texttt{WiFiPenTester} explicitly enforces budget-aware execution, mandatory operator approval, monitor-mode opt-in verification, structured evidence logging, and prompt persistence to support auditability and experimental reproducibility.

This work is guided by the following research questions:
\begin{itemize}
	\item \textbf{RQ1:} To what extent can GenAI assist in selecting viable wireless targets and attack strategies from reconnaissance data under human supervision?
	\item \textbf{RQ2:} How does structured prompt engineering and decision framing influence the accuracy, consistency, and safety of LLM-driven recommendations in wireless PenTesting contexts?
	\item \textbf{RQ3:} What practical limitations arise when deploying LLM-based reasoning within live, time-constrained, and non-deterministic RF environments?
\end{itemize}

To explore these questions, we designed and implemented \texttt{WiFiPenTester} on a Kali Linux testbed equipped with commodity wireless adapters, and tested it  across multiple, controlled wireless environments. The system performs governed monitor-mode validation, passive scanning, GenAI-assisted target prioritisation, controlled deauthentication and handshake capture, and dictionary-based password assessment, while recording detailed execution traces and model interactions. Our results demonstrate that GenAI support can substantially improve target prioritisation accuracy and operational efficiency, while also revealing important sensitivities related to prompt structure, incomplete environmental context, and fluctuating wireless conditions.

This paper makes the following contributions:
\begin{itemize}
	\item \textbf{C1:} a governed GenAI integration approach (or model) for wireless PenTesting that enforces bounded autonomy through explicit human approval checkpoints and a strict separation between LLM reasoning and wireless execution;
	\item \textbf{C2:} demonstrating the feasibility of GenAI-assisted wireless PenTesting under explicit human oversight;
	\item \textbf{C3:} the design and implementation of \texttt{WiFiPenTester} as a modular proof-of-concept (PoC\footnote{In accordance with responsible disclosure principles, only non-sensitive components will be released as open source (\url{https://github.com/DrHaitham/WiFiTesterPP}).}) system that integrates GenAI into IEEE~802.11 reconnaissance and decision support workflows;
	\item \textbf{C4:} a structured prompt-engineering and output-constraining approach for wireless target ranking and feasibility estimation, including deterministic JSON schemas, prompt persistence, and budget-aware token/cost gating for auditable model interactions;
	\item \textbf{C5:} an evidence-centric experimental workflow for reproducible wireless assessment, including session-scoped artefact management that archives scan outputs, normalised metadata, LLM transcripts, executed command traces, and validation records; 
	\item \textbf{C6:} a critical analysis of the system’s design choices, operational benefits, and limitations, including governance trade-offs, model reliability constraints, protocol-dependent feasibility boundaries, and implications for safe deployment of GenAI in wireless security testing; and
	\item \textbf{C7:} an empirical evaluation conducted in authorised and controlled   wireless test environments using commodity hardware and standard toolchains, demonstrating improved target selection efficiency and consistency, and documenting practical failure modes and operational limitations (including constrained WPA3-SAE support).
\end{itemize}

The remainder of this paper is organised as follows. Section~\ref{RelatedWork} reviews related work, while   section~\ref{Background} provides the necessary background on wireless PenTesting and GenAI foundations. Section~\ref{WiFiPenTester_Operation} describes the operational workflow of \texttt{WiFiPenTester}. Section~\ref{WiFiTesterPP_DesignRationale} discusses the key design principles and governance rationale underpinning the system. Section~\ref{PrototypeImplementation} outlines the prototype implementation and experimental setup. Section~\ref{DiscussionAndImplications} analyses the benefits, limitations, and broader implications of the proposed approach. Finally, Section~\ref{ConclusionsAndFurtherResearch} concludes the paper and outlines directions for future work.

\section{Related Work}
\label{RelatedWork}

The application of AI to cybersecurity has received significant research attention, spanning intrusion detection, malware analysis, vulnerability discovery, and offensive security, including ethical hacking. Despite notable progress in automated PenTesting and AI-assisted exploitation~\cite{
	PSTPaper_PenTestModel_2010,
	YaroslavStefinko_ManualAndAutomatedPenetrationTesting_2016,
	2018_AutomatedPenetrationTestingAnOverview_FarahAbuDabaseh,
	2023_GettingGettingPwndbyAI_PenTestingWithLLMs_AndreasHappe,
	2024_ComprehensiveOverviewLLMs_MohammedHassanin,
	2024_PentestGPT_Deng,
	2025_Penterep_WilliLazarov
}, 
 the development of fully autonomous, reliable, and comprehensive PenTesting systems remains an open challenge, particularly in dynamic environments such as wireless networks.

Traditional wireless security assessment tools, including \texttt{Aircrack-ng}\footnote{\url{https://www.aircrack-ng.org/}}~\cite{2022_AircrackNG_ChristopheDevine}, \texttt{Reaver}\footnote{\url{https://www.kali.org/tools/reaver/}}~\cite{2011_WPS_Stefanviehbock}, \texttt{Kismet}\footnote{\url{https://www.kali.org/tools/kismet/}}~\cite{Kismet_MikeKershaw}, and \texttt{Wifite}\footnote{\url{https://www.kali.org/tools/wifite/}}~\cite{2020_wifite2_Deric}, provide some level of automation for scanning, handshake capture, and key recovery.  
However, these systems rely primarily on static heuristics, manual target selection, and operator intuition when determining attack feasibility and prioritisation. They offer limited support for structured reasoning, adaptive strategy selection, or traceable decision processes, and do not incorporate modern AI-based analysis techniques.

Recent work has explored the use of LLMs to support offensive security tasks. PentestGPT~\cite{2024_PentestGPT_Deng} introduced an LLM-powered PenTesting assistant employing modular reasoning, generation, and parsing components within a HITL  workflow. It utilises PenTest Task Trees (PTTs) to preserve context across interactions and mitigate reasoning drift. However, PentestGPT operates primarily as a guided assistant, requiring manual command execution and feedback, thereby limiting the level of automation it can achieve. 

Several   systems have emerged in parallel with our own  work,  which  investigated LLM-driven PenTesting automation, including multi-agent architectures and autonomous exploit orchestration~\cite{2023_GettingGettingPwndbyAI_PenTestingWithLLMs_AndreasHappe,2025_Penterep_WilliLazarov}. These efforts demonstrate the feasibility of LLM-guided attack planning in wired network and application-layer contexts but do not explicitly consider the temporal, probabilistic, and client-dependent characteristics of wireless environments, nor the legal and operational risks associated with uncontrolled active wireless attacks.

Our own prior research examined the role of GenAI in ethical hacking across multiple phases of the PenTesting lifecycle. In~\cite{STM24_UnleashingAIinEthicalHacking}, we proposed a conceptual system for integrating GenAI into  PenTesting workflows. Subsequent studies evaluated GenAI-driven  assistance in  Windows environments~\cite{TechReportUnAIInEH_HC_2024} and Linux platforms~\cite{ITASEC25Paper_AdvancingEthicalHackingWthAIALinux-basedExperimentalStudy_2025}, as well as its application to manual exploitation and privilege escalation~\cite{2024_APracticalExaminationofManualExploitationPrivilegeEscalation_HC}. These works demonstrated that LLMs can enhance analyst productivity, support reasoning under uncertainty, and assist in multi-stage attack planning when properly constrained.

Building on this foundation, \texttt{PenTest++}~\cite{2025_PenTest++_HC_CyBAI,PenTest++_2025pentestelevatingethicalhacking_HaithamChris} introduced a user-centric AI-powered system for automating large portions of the PenTesting workflow while maintaining human oversight. \texttt{PenTest2.0}~\cite{PenTest20_2025autonomousprivesc_HaithamChris,2026_PenTest20_AINA26_HaithamChris} further extended this approach by supporting autonomous, multi-turn privilege escalation driven by LLM reasoning, real-time command execution, structured output parsing, persistent task tracking, and cost-aware governance within a HITL model.

In contrast to prior work, the present study focuses explicitly on the wireless domain and introduces \texttt{WiFiPenTester}, a GenAI-enabled system for intelligent target selection and attack feasibility analysis in IEEE~802.11 environments. The present work extends our own earlier research on GenAI-assisted PenTesting by investigating how LLMs can be integrated into wireless ethical hacking workflows. Rather than pursuing full automation of wireless attacks, the system confines LLM usage to the reconnaissance and decision-support stages, while enforcing explicit human operator approval for all active operations, thereby supporting reproducibility, cost control, auditability, and responsible deployment.
Indeed, wireless assessment presents additional challenges: partial observability due to RF propagation, time-varying client behaviour, and the potential disruption of legitimate users during active interactions. Such  properties amplify the risks of hallucinated or overconfident LLM recommendations, thereby making human governance essential.

More fundamentally, this represents a shift in both technical focus and system design compared to our earlier GenAI-assisted PenTesting platforms.  
Previous systems such as \texttt{PenTest++}~\cite{2025_PenTest++_HC_CyBAI,PenTest++_2025pentestelevatingethicalhacking_HaithamChris} and \texttt{PenTest2.0}~\cite{PenTest20_2025autonomousprivesc_HaithamChris,2026_PenTest20_AINA26_HaithamChris} operate in comparatively stable, host-centric \emph{wired-network} environments supporting initial exploitation and post-compromise activities, where the LLM reasons over persistent system state and automated actions affect only the target machine. 
In contrast, \texttt{WiFiPenTester} addresses   pre-access wireless PenTesting, where feasibility is inherently probabilistic, time-sensitive, and jointly determined by protocol configuration and volatile physical-layer conditions (e.g., signal quality, interference, and client mobility). Moreover, wireless actions can impact  third-party devices and live services, elevating both operational risk and ethical exposure. These differences motivate a governance-first integration of GenAI, in which bounded autonomy and strict separation between reasoning and execution are not optional safeguards but core architectural requirements.

\section{Background}
\label{Background}

We now provide  the technical and conceptual foundations required to understand the design and evaluation of \texttt{WiFiPenTester}. 

\subsection{IEEE~802.11 Security Fundamentals}
\label{IEEE~802.11 Security Fundamentals}

Wireless local area networks (WLANs) based on IEEE~802.11 enable mobility but expose communication to both passive eavesdropping and active interference. Security mechanisms have evolved substantially over the last two decades, progressing from the fundamentally flawed Wired Equivalent Privacy (WEP)~\cite{2006_WEP_AndreaBittau} protocol to Wi-Fi Protected Access (WPA), WPA2~\cite{2009_ASurveyOnWirelessSecurityProtocols_ArashHabibi}, and more recently WPA3~\cite{2025_WPA3_WiFiAlliance,2020_DragonbloodWPA3_MathyVanhoef}, which is discussed further in Section \ref{sec:wpa3-sae-background}.

WEP relies  on the RC4 stream cipher with short initialisation vectors, enabling practical key recovery through statistical analysis of IV reuse, rendering the protocol obsolete shortly after deployment~\cite{2006_WEP_AndreaBittau}. WPA introduced TKIP (Temporal Key Integrity Protocol) as an interim mitigation, while WPA2 standardised AES-CCMP for confidentiality and integrity protection~\cite{2015_PlaintextRecoveryAttacksAgainstWPA/TKIP_KennethPaterson}. The security of WPA/WPA2-PSK (Pre-Shared Key) networks  depends on the entropy of the shared passphrase and on client behaviour, since captured authentication material can support offline password guessing.

In parallel, 802.11 management frames (e.g., for deauthentication and disassociation) were historically unauthenticated, enabling denial-of-service (DoS) and handshake-forcing attacks~\cite{2003_802.11DoSAttacks_JohnBellardo}. Although 802.11w introduced Management Frame Protection (MFP), real-world deployment remains inconsistent, and many environments still permit unauthenticated deauthentication traffic~\cite{2003_802.11DoSAttacks_JohnBellardo}.

\subsection{WPA3 and SAE Attack Surface}
\label{sec:wpa3-sae-background}

WPA3~\cite{2025_WPA3_WiFiAlliance} was introduced to address key weaknesses of WPA2, particularly the susceptibility of WPA2-PSK to offline dictionary attacks after handshake capture. In WPA3-Personal, the PSK handshake is replaced by Simultaneous Authentication of Equals (SAE), a password-authenticated key exchange (PAKE) based on a Dragonfly-style construction~\cite{2015_RFC7664_Harkins}. SAE is intended to provide \emph{offline dictionary attack resistance}: an observer who captures authentication traffic should not be able to validate password guesses offline without interacting with the AP.

Nevertheless, WPA3/SAE introduces a distinct and practically relevant attack surface. First, SAE reduces offline guessing but does not eliminate \emph{online} guessing attempts; practical resilience depends on correct anti-clogging behaviour, rate limiting, and implementation robustness. Second, many real deployments use transitional configurations (e.g., “WPA2/WPA3 mixed mode”) to support legacy clients. Mixed-mode operation can, in practice, reintroduce downgrade opportunities or client-driven fallbacks, depending on AP configuration and client capability negotiation \cite{2020_DragonbloodWPA3_MathyVanhoef}.

A well-known example is the “Dragonblood” family of attacks, which demonstrated that certain design choices and implementation flaws can enable downgrade attacks, side-channel leakage, or DoS in some WPA3 deployments~\cite{2020_DragonbloodWPA3_MathyVanhoef}. Although patches and configuration hardening mitigate many practical risks, these results highlight that WPA3’s real-world security hinges on careful configuration and correct implementations, not merely the presence of SAE.

For ethical hacking, these properties shift assessment emphasis compared to WPA2-PSK. Instead of assuming that capture enables offline cracking, WPA3~\cite{2023_WPA3-ASystematicLiteratureReview_AsmaaHalbouni,2019_ReviewOfWEP-WPA2-WPA3_IndiraReddy} testing often focuses on: (i) configuration verification (e.g., SAE-only where feasible); (ii) assessing downgrade resilience in transitional modes; (iii) verifying MFP and client behaviour; and (iv) identifying operational weaknesses such as weak passphrases combined with insufficient rate limiting. Accordingly, a governed wireless assessment system should treat WPA3 targets as a distinct class, with feasibility estimation driven by protocol mode, observed client capabilities, and configuration posture rather than by WPA2-style handshake-cracking assumptions.

\subsection{Wireless Penetration Testing Workflow}
\label{Wireless Penetration Testing Workflow}

Wireless PenTesting evaluates the resilience of WLAN deployments. 
In practice, a typical assessment proceeds through several tightly coupled stages, beginning with preparation of the wireless interface in monitor mode and configuration of channel hopping or fixed-channel capture, followed by passive reconnaissance to collect beacon frames and probe responses for enumerating APs, operating channels, advertised security capabilities, and signal characteristics. Subsequent client observation aims to identify associated stations and short-term activity levels in order to estimate the feasibility of time-sensitive operations, such as authentication capture. Where permitted by scope and risk constraints, controlled active interaction may then be employed, for example through the transmission of deauthentication frames to stimulate reconnection events, enabling the capture of authentication handshakes for offline analysis. Finally, resilience testing may be conducted against captured material using dictionary-based verification or broader password auditing techniques when explicitly authorised. Although these stages are conceptually sequential, in real deployments they often overlap and require continuous reassessment as environmental conditions and client behaviour evolve.

\subsection{Limitations of Existing Wireless  Tools}
\label{Limitations of Existing Automation Frameworks}

As previously discussed in Section~\ref{RelatedWork}, established toolchains, e.g. \texttt{aircrack-ng}, \texttt{airodump-ng}, \texttt{aireplay-ng}, \texttt{Reaver}, and \texttt{Wifite}, automate substantial portions of this process, particularly low-level packet capture, frame injection, and cryptographic verification. However, the most important aspects of the workflow remain judgement-intensive. Decisions concerning which network to prioritise, whether sufficient client activity exists to justify active interaction, which channel and time window to select, and how to balance operational impact against engagement objectives are typically left to the operator. Integrated automation frameworks such as \texttt{Wifite} attempt to streamline execution by managing multiple tools, yet their decision logic is predominantly heuristic-based, relying on simple indicators such as signal strength, encryption type, Wi-Fi Protected Setup (WPS) availability, or fixed thresholds. Such heuristics often perform poorly in dense or dynamic environments where many networks coexist, RF conditions fluctuate, and client behaviour is inherently non-deterministic.

Beyond these operational limitations, existing wireless automation systems, notably \texttt{Wifite}, provide little support for GenAI-assisted reasoning and exhibit   shortcomings that constrain their suitability for governed deployment and systematic evaluation. Our preliminary lab-based tests indicate that they offer little   explanation of \emph{why} a particular target is recommended, resulting in limited transparency into the decision process underlying automated choices. Intermediate decisions and contextual factors are not consistently recorded in structured form, which limits both reproducibility and auditability and complicates post-engagement review and research replication. As wireless environments become larger and more complex, these issues increase analyst workload and make oversight harder, which motivates the need for AI-driven decision support under human control.

 \subsection{Received Signal Strength Indicator (RSSI)}
 \label{Received Signal Strength Indicator (RSSI)}
 
 RSSI is a widely used metric that quantifies the power level of a radio signal as observed by a receiving wireless interface~\cite{2025_WiFiSignalStrengthGuide_DannyMareco}. Although it is not standardised to a single absolute scale across chipset vendors, RSSI is commonly expressed in decibel-milliwatts (dBm) and serves as a practical indicator of link quality, propagation conditions, and physical proximity between a station and an AP. 
 
 In wireless security assessment, RSSI plays a central role in estimating the feasibility and reliability of packet capture, client interaction, and handshake acquisition, as weak signal conditions increase frame loss, timing instability, and decoding errors. Consequently,  PenTesting tools routinely use RSSI thresholds to prioritise targets, select channels, and adapt capture strategies.
 
 Table~\ref{tab:rssi_interpretation} summarises typical interpretations of RSSI ranges in IEEE~802.11 environments and their operational implications for wireless attacks and monitoring activities. In \texttt{WiFiPenTester}, RSSI is treated as a key feature during structured metadata aggregation and GenAI-assisted target ranking, enabling the system to favour networks that are not only cryptographically weaker but also operationally reachable under realistic RF conditions.

\begin{table}[h]
	\centering
	\caption{Typical interpretation of RSSI values in IEEE 802.11 networks}
	\label{tab:rssi_interpretation}
	
	\renewcommand{\arraystretch}{1.25}
	\setlength{\tabcolsep}{6pt}
	
	\rowcolors{3}{gray!8}{white}
	
	\begin{tabular}{c c l}
		\toprule
		\toprule
		\textbf{RSSI Range (dBm)} & \textbf{Signal Quality} & \textbf{Operational Implication} \\
		\midrule
		$\geq -30$ & Excellent &
		\makecell[l]{Very strong signal; stable communication\\ and reliable packet capture expected.} \\
		
		From $-31$  to $-50$ & Good &
		\makecell[l]{Suitable for sustained monitoring\\ and handshake capture.} \\
		
		From   $-51$ to  $-70$ & Weak but usable &
		\makecell[l]{Increased frame loss possible; active\\ interaction may be unreliable.} \\
		
		$\leq -85$ & Poor / Unusable &
		\makecell[l]{High packet loss; handshake capture and\\ active attacks are unlikely to succeed.} \\
		
		\bottomrule
	\end{tabular}
\end{table}

\section{WiFiPenTester Operation}
\label{WiFiPenTester_Operation}
\texttt{WiFiPenTester} is   a governed,  system for GenAI-assisted wireless PenTesting that augments traditional WiFite-style workflows with structured decision support, human oversight, and full execution traceability. The system is intended to be used in fully authorised and controlled assessment environments. The system  operates as follows (see Fig.~\ref{WiFiTesterPPArchitecture2}).

\begin{enumerate}
	
	\item \textbf{System readiness validation:}  
	The system verifies wireless interface availability, driver compatibility, regulatory domain configuration, and toolchain dependencies before any RF operation is permitted.
	
	\item \textbf{Monitor-mode activation:}  
	The selected wireless interface is transitioned into monitor mode only after explicit user confirmation, ensuring that potentially disruptive behaviour is always intentional and authorised.
	
	\item \textbf{Passive reconnaissance and client observation:}  
	Beacon frames and probe responses are collected to enumerate nearby APs, encryption schemes, authentication modes, channels, signal strength indicators, and advertised capabilities. 
	Simultaneously, associated stations and traffic rates are also monitored to estimate temporal feasibility of handshake capture and the likelihood of successful active interaction.
	
	\item \textbf{Structured metadata aggregation:}  
	\label{Step-StructuredMetadataAggregation} 
	All observed network information is normalised into an internal structured representation that captures the essential properties of each discovered wireless network. This includes the Basic Service Set Identifier (BSSID) and the Extended Service Set Identifier (ESSID), the encryption and authentication configuration (e.g., WEB,  WPA2, WPA3, mixed-mode operation, or  WPS), the operating channel and the RSSI (see Section \ref{Received Signal Strength Indicator (RSSI)}), the number of associated client stations, short-term traffic activity levels, and relevant protocol features such as the presence of MFP. 
	 
	\item \textbf{Prompt construction:}  
	\label{Step-GenAIPromptConstruction} 
	The system injects the aggregated metadata into a predefined prompt template that:
	
	\begin{itemize}
		\item explicitly assigns the LLM an expert role (``a seasoned wireless penetration tester'') to frame its reasoning style and domain assumptions;
		\item restricts the LLM to advisory reasoning only (i.e., target ranking and feasibility analysis, without triggering actions);
		\item enforces a deterministic, structured JSON output schema with fixed fields for scores, justification, confidence, and recommended targets;
		\item defines task-specific scoring criteria and interpretation guidelines to reduce ambiguity and improve consistency of model recommendations in dense wireless environments; 
		\item provides session context (timestamps, session identifier, tool name, and assessment stage) to ensure traceability and reproducibility across runs; and
		\item embeds the authorised testing context and scope of engagement, and encodes legal, ethical, and operational constraints.  
	\end{itemize}
	
	This design ensures that the LLM reasons over concrete, structured wireless observations rather than free-form descriptions, while remaining bounded by explicit governance, auditability, and human oversight requirements.
	
	\item \textbf{Budget estimation and human approval:}  
	Token count and estimated API cost are computed and presented to the user together with the full prompt. Only after explicit approval is the prompt submitted to the LLM.
	
	\item \textbf{LLM-based target ranking and feasibility estimation:}  
	The LLM returns a structured response containing:
	
	\begin{itemize}
		\item ranked candidate targets;
		\item feasibility scores;
		\item justifications grounded in observed metadata; and
		\item qualitative risk indicators.
	\end{itemize}
	
	The raw response is persisted verbatim, and a validated JSON parse is generated for internal use. 
	
	\item \textbf{HITL target selection:}  
	The operator reviews both the model’s recommendation and the raw scan data before selecting the actual target.   This step enforces bounded autonomy: GenAI advises, but never decides.
	
	\item \textbf{Controlled active interaction:}  
	Only after human selection does the system  proceed to channel locking, optional deauthentication, and handshake capture, using conventional tools under strict logging and attribution.
	
\item \textbf{Protocol-aware and key-strength assessment:}  
\label{step:handshake-validation} 
Captured authentication exchanges are first validated for structural completeness and protocol correctness to ensure that subsequent analysis is meaningful and reproducible. The validation procedure is protocol-aware and is designed to proceed  as follows.

\begin{itemize}
	\item \textbf{WEP:}  
	Captured IV-rich traffic traces are inspected to confirm sufficient packet volume and entropy for statistical key recovery attacks. When adequate data is available, automated key recovery may be attempted using standard FMS\footnote{FMS\cite{2001_RC4_FMS} refers to the Fluhrer–Mantin–Shamir attack, a statistical RC4 key-recovery attack that exploits weak IVs in WEP.}/Korek-style techniques to evaluate whether the network remains vulnerable to legacy cryptographic weaknesses.
	
	\item \textbf{WPA/WPA2-PSK:}  
	Four-way handshakes are verified for completeness and temporal consistency. When a valid handshake is confirmed, dictionary-based assessment may be carried out using an operator-defined wordlist
	 to evaluate passphrase resilience against offline guessing attacks. The outcome is recorded as either \emph{recovered}, \emph{not recovered}, or \emph{insufficient evidence}, together with timing and attempt statistics.
	
	\item \textbf{WPA3-SAE:}  
	As offline key recovery is computationally infeasible by design, \texttt{WiFiPenTester} instead evaluates the security posture of the deployment, including:
	\begin{itemize}
		\item confirmation of SAE usage versus transitional (mixed WPA2/WPA3) modes;
		\item detection of downgrade exposure;
		\item assessment of MFP (802.11w); and
		\item identification of configuration patterns associated with known weaknesses (e.g., Dragonblood-class design limitations).
	\end{itemize}
\end{itemize}

Rather than relying on heuristic success or failure alone, the system records protocol-specific evidence describing \emph{why} an assessment outcome was reached. This ensures that the system reasons over concrete wireless observations rather than free-form descriptions, and that all conclusions remain technically auditable.

	\item \textbf{Evidence consolidation:}  
	All operational artefacts are collected, including:
	
	\begin{itemize}
		\item raw scan outputs;
		\item structured network metadata;
		\item LLM prompts and responses;
		\item token usage and costs;
		\item executed commands \& timestamps; 
		\item handshake status; and
		\item assessment outcomes.
	\end{itemize}
	
	\item \textbf{GenAI-assisted report generation:}  
	\label{GenAI-assisted report generation:}
	At the conclusion of the assessment, \texttt{WiFiPenTester} constructs a final reporting prompt containing all relevant technical facts, observations, and outcomes, and submits it to the LLM to generate a structured PenTesting report in JSON format.

	\item \textbf{Termination and archival:}  
	The workflow terminates upon completion of reporting, timeout, or user intervention. All artefacts are archived to support reproducibility, auditability, and subsequent research analysis.
	
\end{enumerate}

\section{Design Principles and Architectural Rationale}
\label{WiFiTesterPP_DesignRationale}
This section outlines the key design principles underpinning \texttt{WiFiPenTester}.
\subsection{Human-Governed Active Interaction and Operational Ordering}

The ordering of steps in Section~\ref{WiFiPenTester_Operation} is itself a design decision. Passive reconnaissance and client observation are conducted first, providing a factual basis for feasibility reasoning and limiting unnecessary transmissions. Active interaction is placed only after target selection and explicit user approval. This ordering ensures that disruptive actions are justified by observed conditions (e.g., the presence of clients, sufficient signal quality, and plausible handshake capture opportunities). It also reflects a practical constraint of wireless assessments: feasibility is time-dependent and context-sensitive, and therefore cannot be reduced to static heuristics. By design, \texttt{WiFiPenTester} allows GenAI to assist with prioritisation and feasibility estimation, but it maintains explicit operator approval of any operation that might affect availability.  

\subsection{Bounded Autonomy as a Hard Requirement}

Wireless PenTesting operates in a shared, RF environment where active actions may affect third-party devices and production services. In contrast to post-exploitation tasks performed within an assumed-breach host context (as, e.g., in PenTest2.0~\cite{PenTest20_2025autonomousprivesc_HaithamChris,2026_PenTest20_AINA26_HaithamChris}), wireless operations such as deauthentication and channel locking can cause observable disruption and may carry heightened legal and ethical exposure. For this reason, \texttt{WiFiPenTester} adopts bounded autonomy as a non-negotiable requirement: GenAI provides recommendations and structured reasoning, but the operator retains sole authority over decisions and all active transmissions. Concretely, the workflow enforces explicit user approval before each LLM invocation and before any active wireless action, ensuring that model output cannot directly trigger potentially disruptive behaviour. This decision is motivated by the need for operational safety, scope compliance, and accountable use of offensive capabilities in real environments.

\subsection{Prompt Design as Rules of Engagement}

As described in Section~\ref{WiFiPenTester_Operation} (Step~\ref{Step-GenAIPromptConstruction}), the behaviour of a general-purpose LLM is highly sensitive to the structure, context, and constraints encoded in its input prompt \cite{2022_InteractiveAndVisualPromptEngineeringWithLLMs_HendrikStrobelt,2023_ConversingWithCopilotPromptEngineering_PaulDenny,2022_TrainingLanguageModelsToFollowInstructionsWithHumanFeedback_LongOuyang,2023_PreTrainPromptPredict_PengfeiLiu}. In \texttt{WiFiPenTester}, prompt construction is therefore treated as an explicit mechanism for defining the system’s \emph{rules of engagement}, rather than as a simple formatting step.

In particular, the model is explicitly assigned the role of a ``seasoned wireless penetration tester'' in order to anchor its reasoning in realistic attacker and assessor mental models, domain-specific terminology, and protocol-aware threat assessment practices. Without such role specification, the model outputs could tend to drift towards generic security advice, inconsistent scoring, or overly conservative recommendations that are misaligned with practical wireless assessment objectives.

The prompt further constrains the model to an advisory function only, restricting it to target ranking and feasibility analysis while prohibiting action generation. This separation preserves human control over all active operations and prevents the LLM from implicitly assuming operational authority. Structured JSON output schemas, fixed scoring criteria, and interpretation guidelines are additionally enforced to reduce ambiguity, improve comparability across sessions, and enable systematic downstream processing and auditing.

Finally, session metadata and explicit legal and ethical constraints are embedded directly within the prompt to ensure traceability, reproducibility, and continuous reinforcement of the authorised testing context. Collectively, these design choices transform the prompt from an informal query into a formally governed interface that bounds model behaviour, aligns recommendations with professional assessment standards, and supports safe integration of GenAI within security-critical workflows.

\subsection{Use of Chain-of-Thought (CoT) Reasoning}
\label{sec:cot-rationale}

Recent work~\cite{2022_ChainOfThoughtPrompting_JasonWei,2023_TreeOfThoughts_ShunyuYao} has shown that CoT prompting can significantly improve the quality and consistency of LLM reasoning by encouraging models to decompose complex decisions into intermediate analytical steps. In the context of wireless PenTesting, such decomposition is particularly valuable, as target prioritisation and feasibility estimation depend on multiple interacting factors, including protocol configuration, signal quality, client activity, and environmental stability. \texttt{WiFiPenTester} therefore adopts CoT reasoning in a constrained and explicitly governed manner: the system’s prompt templates instruct the LLM to reason step-by-step over the structured scan metadata before producing its final recommendation, thereby improving coherence and reducing shallow heuristic decisions, while remaining strictly limited to advisory analysis only and prohibited from generating commands, triggering actions, or altering execution flow.

\subsection{Reasoning over Structured Observations, not Free-Form Logs}

A central objective of \texttt{WiFiPenTester} is to ensure that LLM reasoning is grounded in verifiable observations rather than ambiguous or stylistically variable tool output. Raw scan logs and packet-level artefacts are often noisy, vendor-dependent, and difficult to interpret consistently, both for humans and for LLMs. Accordingly, the system normalises reconnaissance results into a structured internal representation prior to prompt construction. This representation includes only the fields required for decision support (e.g., security mode, channel, RSSI, client activity indicators, and relevant protocol features). The prompt is then constructed by injecting these structured facts into a constrained template. As a result, the LLM is compelled to reason over concrete wireless observations rather than free-form descriptions, improving robustness, reducing hallucination risk, and supporting deterministic downstream parsing of the model’s structured JSON response.

\subsection{Strict Separation between Reasoning and Execution}

\texttt{WiFiPenTester} deliberately separates GenAI-driven reasoning from operational execution. All wireless actions are performed by deterministic local tooling (e.g., \texttt{airodump-ng}, \texttt{aireplay-ng}, \texttt{aircrack-ng}), while the LLM is confined to advisory functions such as target ranking, feasibility estimation, qualitative risk assessment, and final report generation. This separation mitigates prompt-injection and instruction-following risks by preventing model output from becoming executable control flow. It also preserves experimental repeatability: the execution path remains attributable to known tools and parameters, while the model’s role is clearly scoped to analysis and documentation. In practice, this design is essential for building a system that can be evaluated rigorously, since the effects of LLM variability are isolated from the mechanics of radio operations.

\subsection{Protocol-Aware Validation and Assessment Strategy}

Wireless security assessment cannot be treated as a single uniform procedure, as the available evidence, attack surfaces, and feasible analytical techniques differ fundamentally across WEP, WPA/WPA2, and WPA3 deployments. Accordingly, \texttt{WiFiPenTester} is designed around a protocol-aware validation and assessment strategy that adapts its reasoning and evaluation objectives to the underlying security mechanism in use.

Concretely, as described in Section~\ref{WiFiPenTester_Operation} (Step~\ref{step:handshake-validation}), the system distinguishes between protocol-specific assessment goals for \texttt{WEP}, \texttt{WPA/WPA2-PSK}, and \texttt{WPA3-SAE}, applying different validation criteria and security posture evaluations depending on the protocol in operation.

This protocol-aware treatment prevents misleading generalisations across heterogeneous security mechanisms and enables the system to report wireless security posture in a manner that accurately reflects cryptographic design intent, operational constraints, and realistic attacker capabilities.

\subsection{Privacy-Preserving GenAI Integration}
\label{Privacy-Preserving GenAI Integration}

As introduced in Section~\ref{WiFiPenTester_Operation} (Step~\ref{GenAI-assisted report generation:}),  \texttt{WiFiPenTester} employs GenAI   both for decision support and for assisting in the generation of structured assessment reports.
This improves usability and supports consistent documentation; however, it introduces privacy and confidentiality considerations. The reporting design therefore follows a strict data-minimisation policy. Raw packet captures, cryptographic material, and client identifiers are not submitted to the LLM. Most importantly, cracked credentials are never transmitted in plaintext. Instead, the report-generation prompt includes only factual outcomes (e.g., ``passphrase recovered''
 or ``dictionary attempt unsuccessful'') and may include masked representations where necessary for narrative clarity. This design preserves confidentiality while still enabling the LLM to produce a useful, structured PenTesting report. It also aligns with recognised principles of data minimisation and purpose limitation, and supports deployment in environments with strict governance requirements.

In addition, the architecture does not mandate the use of cloud-based LLMs. To further reduce the risk of data exposure, \texttt{WiFiPenTester} can be configured to operate with locally deployed LLMs (e.g., \texttt{llama.cpp\footnote{\url{https://github.com/ggml-org/llama.cpp}}}~\cite{2023_LlamaCppLocalInference_GeorgiGerganov,2023_LLaMA_OpenEfficientFoundationLanguageModels_HugoTouvron}, \texttt{Ollama\footnote{\url{https://ollama.com/}}})  executed entirely within the assessment environment. This allows all prompts and generated reports to remain on the assessor’s infrastructure, eliminating third-party data transfer and supporting compliance with strict organisational, regulatory, or air-gapped operational constraints. Such deployment is particularly relevant for governmental and critical-infrastructure environments where external data sharing is generally  prohibited.

\subsection{Cost Awareness as an Explicit System Constraint}

Unlike conventional security tooling, LLM usage incurs direct financial cost and can exhibit unbounded prompt growth if not constrained. \texttt{WiFiPenTester} therefore treats cost awareness as an explicit operational constraint. Before every LLM invocation, the system computes token counts and estimates monetary cost, presenting these to the operator for approval.

This mechanism prevents unexpected expenditure, discourages prompt bloat, and provides consistent telemetry for cost analysis in experimental evaluation. Importantly, it also reinforces the bounded-autonomy model: the operator remains accountable not only for disruptive actions but also for resource consumption and external service usage, which is especially important in budget-sensitive or resource-constrained environments.

This concern is not merely theoretical. As observed during the early development and internal evaluation of \texttt{PenTest2.0}~\cite{PenTest20_2025autonomousprivesc_HaithamChris,2026_PenTest20_AINA26_HaithamChris} , the absence of an explicit cost-control checkpoint led to unintended prompt inflation, rapidly exhausting available API credits and prematurely halting experimental runs. Based on this experience, the manual approval mechanism was introduced as a deliberate design choice and subsequently adopted in \texttt{WiFiPenTester}.

\subsection{Evidence-Centric Execution for Reproducible Research}

A recurring challenge in wireless experimentation is environmental variability and the difficulty of reproducing conditions across runs. \texttt{WiFiPenTester} addresses this by treating each execution as an evidence-producing experiment. The system persistently logs scan outputs, derived structured metadata, prompts and responses, token usage, executed commands, and assessment outcomes. This evidence-centric approach supports verifiability, enables post-hoc failure analysis, and allows results to be contextualised against environmental observations rather than interpreted as isolated success or failure. In combination, the preceding design decisions ensure that \texttt{WiFiPenTester} can be studied scientifically: decisions are attributable, model interactions are auditable, and operational actions are governed by explicit human approval.

\subsection{On LLM Safety Mechanisms and Model Substitution}

Some LLMs employ built-in safety and policy enforcement mechanisms that may cause them to refuse or partially redact responses to prompts related to sensitive or offensive security scenarios, including PenTesting and exploit development. Although such behaviour was not encountered in our PoC experiments, it can represent  a practical consideration for GenAI-assisted security tooling.

In practice, this limitation can often be mitigated through model substitution, as different providers and model families exhibit varying degrees of strictness, calibration, and policy interpretation. A system architecture that abstracts the LLM interface therefore provides resilience against overly restrictive behaviour by allowing assessors to switch between models without modifying the surrounding workflow.

Beyond provider diversity, recent research~\cite{2023_JailbrokenSafetyTrainingFails_AlexanderWei,2023_UniversalTransferableAdversarialAttacks_AndyZou,2025_BypassingLLMGuardrails_WilliamHackett}
 has demonstrated that LLM safety mechanisms are not absolute and can be circumvented under certain conditions. Prior studies have documented a range of general classes of techniques, including prompt re-framing, multi-turn context manipulation, role-based conditioning, and indirect instruction encoding, which can lead models to generate outputs that bypass intended safety constraints. These findings highlight that safety enforcement in contemporary LLMs remains probabilistic and model-dependent rather than formally guaranteed.

From a defensive and system-design perspective, these observations reinforce two points. First, safety controls should not be treated as a substitute for explicit governance and human oversight within security-sensitive applications. Second, the ability to select and replace models should be considered a practical robustness feature rather than a circumvention strategy, enabling continued operation in environments where particular providers impose restrictive policies or experience unpredictable refusal behaviour.

\section{Prototype Implementation}
\label{PrototypeImplementation}
We now describe  a PoC prototype,  developed to empirically evaluate the proposed system architecture.

\subsection{Experimental Setup}
\label{WiFiImplementationSetup}

\texttt{WiFiPenTester} is implemented in Python~3, selected for its rapid prototyping capabilities, mature ecosystem of networking and wireless-security libraries, and seamless integration with modern LLM APIs. All experiments were conducted using VirtualBox~7 on a physical host system running Windows~11 (Lenovo laptop, Intel Core Ultra~7 CPU, 32\,GB RAM). The experimental environment consisted of a Kali Linux virtual machine (VM) hosting \texttt{WiFiPenTester} and a collection of external APs deployed for controlled testing.

Since VirtualBox virtualises wireless interfaces as Ethernet devices and does not expose raw IEEE~802.11 frame access to guest operating systems (OSs), and because the built-in wireless adapter of the host laptop lacks support for monitor mode and packet injection even in non-virtualised environments, an external USB WiFi adapter was required. Specifically, a MediaTek-based USB 802.11n adapter, supporting both monitor mode and packet injection, was attached to the Kali VM via USB passthrough\footnote{MediaTek MT7601U chipset (USB vendor ID: 0x148f, product ID: 0x7601).}. This configuration enabled full passive capture and active interaction, including deauthentication and handshake acquisition, as required by the system. Successful operation in monitor mode was verified at runtime (e.g., \texttt{wlan0} operating in \emph{Mode: Monitor}), ensuring that low-level IEEE~802.11 frame capture and injection were supported throughout the experiments.

For GenAI integration, this prototype employs OpenAI’s \texttt{gpt-4o-mini} model via the official API, chosen for its favourable cost–performance trade-off and stable structured-output behaviour. While the system supports alternative models (e.g., \texttt{o3}, \texttt{gpt-4.1}, \texttt{gpt-5}), these were not used in the reported experiments due to increased token cost and variability. All API interactions are mediated through a dedicated connector module that enforces cost estimation, prompt validation, and user approval prior to submission.

Other alternative LLM providers could also be integrated, including Anthropic’s Claude\footnote{\url{https://www.anthropic.com/claude}}, Google DeepMind’s Gemini\footnote{\url{https://deepmind.google/technologies/gemini/}}, Mistral AI\footnote{\url{https://mistral.ai/}}, and Cohere\footnote{\url{https://cohere.com/}}. In addition, self-hosted open-weight models such as Meta’s LLaMA\footnote{\url{https://ai.meta.com/llama/}} or Qwen\footnote{\url{https://qwenlm.github.io/}} could be deployed locally using lightweight inference frameworks, e.g.\   llama.cpp\footnote{\url{https://github.com/ggerganov/llama.cpp}} or Ollama\footnote{\url{https://ollama.com/}}. 

This flexibility is enabled by the system’s modular LLM connector layer, which abstracts provider-specific APIs and enforces uniform interfaces for prompt construction, cost estimation, validation, and response parsing, thereby supporting portability and experimental reproducibility across different deployment configurations.

\subsection{PoC Implementation}
\label{WiFiPoCImplementation}

The prototype of \texttt{WiFiPenTester} realises a governed, single-session workflow for GenAI-assisted wireless assessment, combining passive reconnaissance, structured metadata analysis,  HITL decision making, and controlled active interaction. The operational logic is implemented as a modular, command-line application executed on a Kali Linux VM and proceeds as follows.

\begin{figure}[h]
	\centering
	\includegraphics[width=1\linewidth]{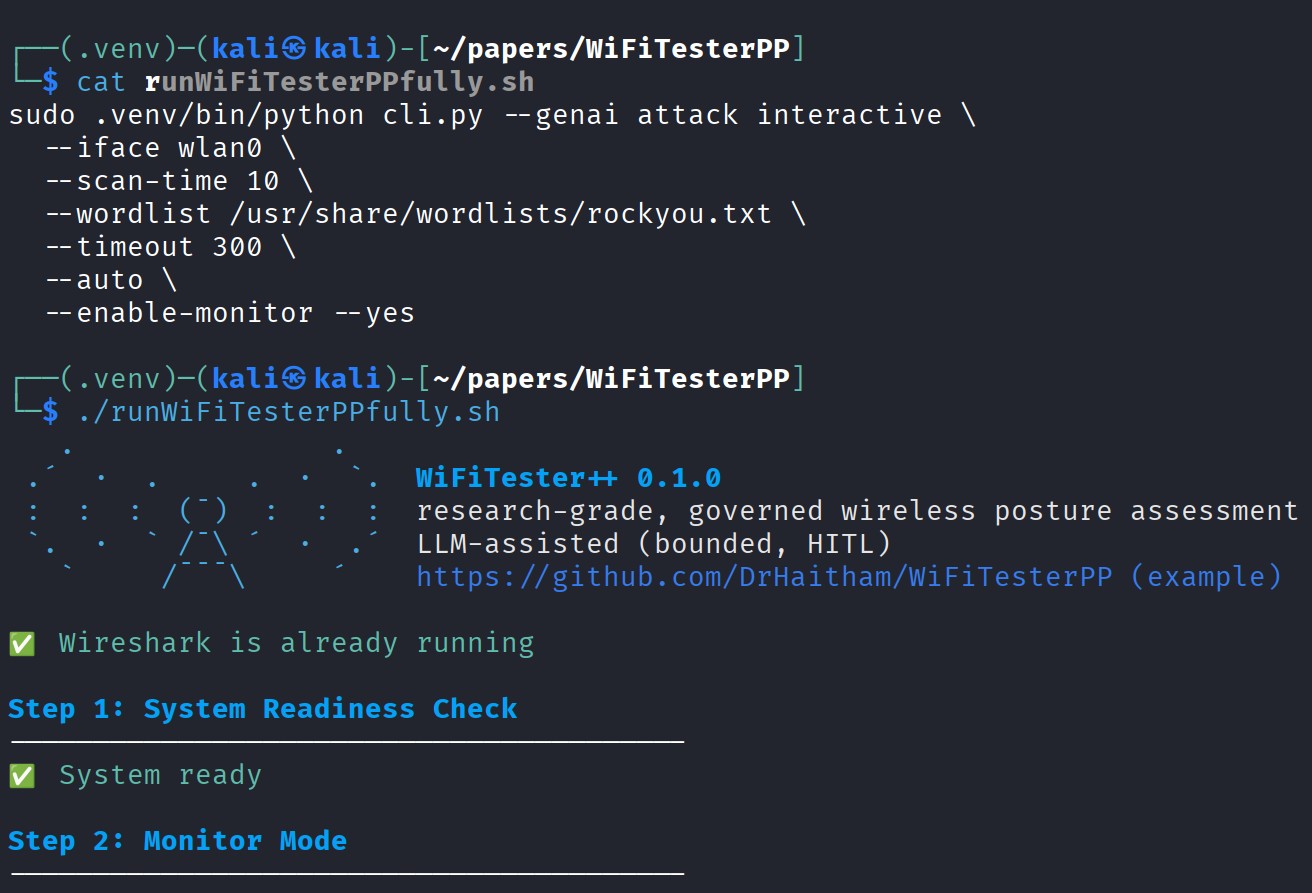}
	\caption{System initialisation and environment validation.}
	\label{fig:imgfirstsuccessrunpart1}
\end{figure}

\begin{figure}[h]
	\centering
	\includegraphics[width=1\linewidth]{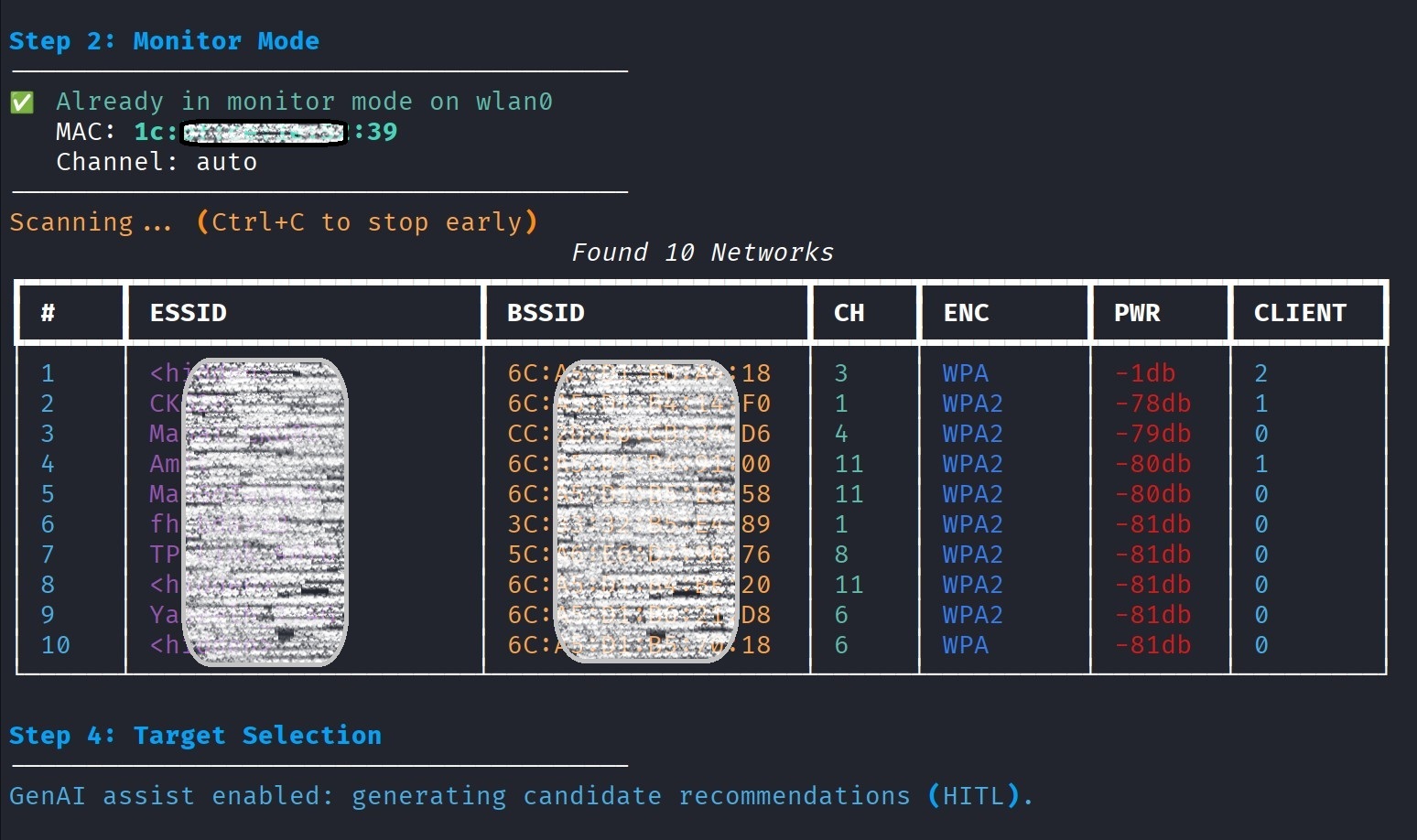}
	\caption{Passive reconnaissance results showing discovered APs and protocol metadata.}
	\label{fig:imgfirstsuccessrunpart2}
\end{figure}

\begin{figure}[h]
	\centering
	\includegraphics[width=1\linewidth]{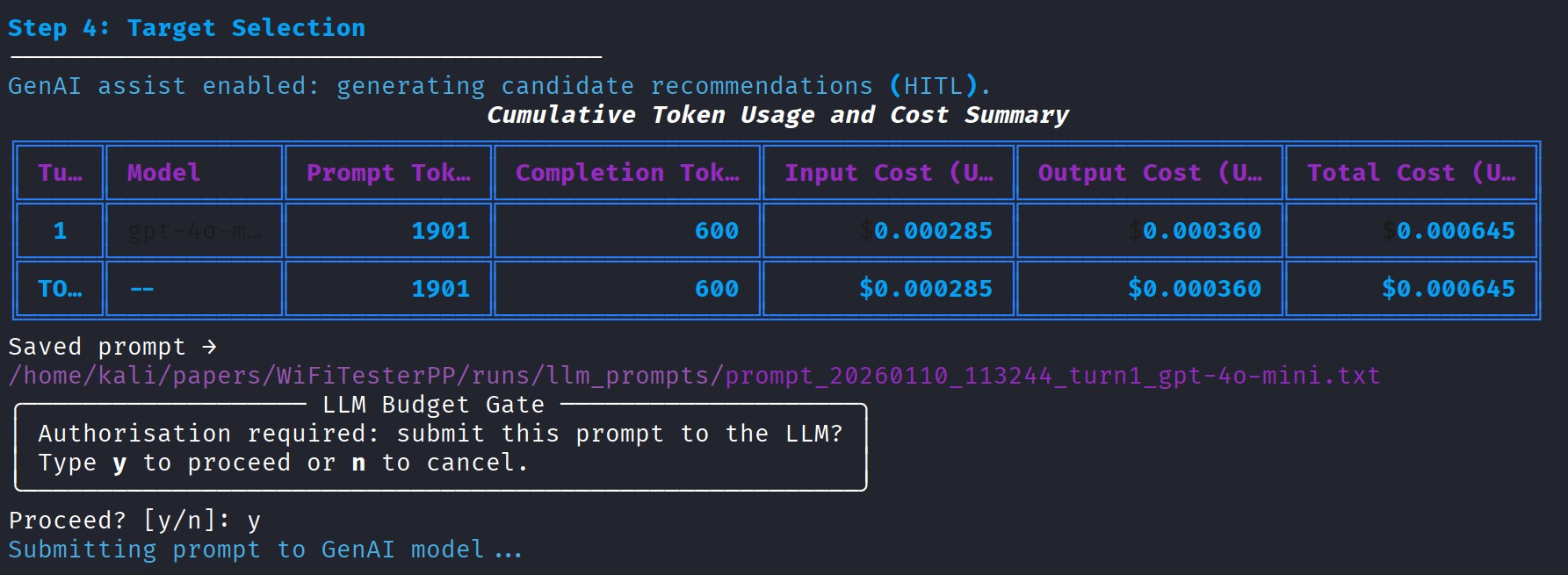}
	\caption{Prompt token statistics and API cost estimation prior to LLM submission.}
	\label{fig:imgfirstsuccessrunpart3}
\end{figure}

\begin{figure}[h]
	\centering
	\includegraphics[width=1\linewidth]{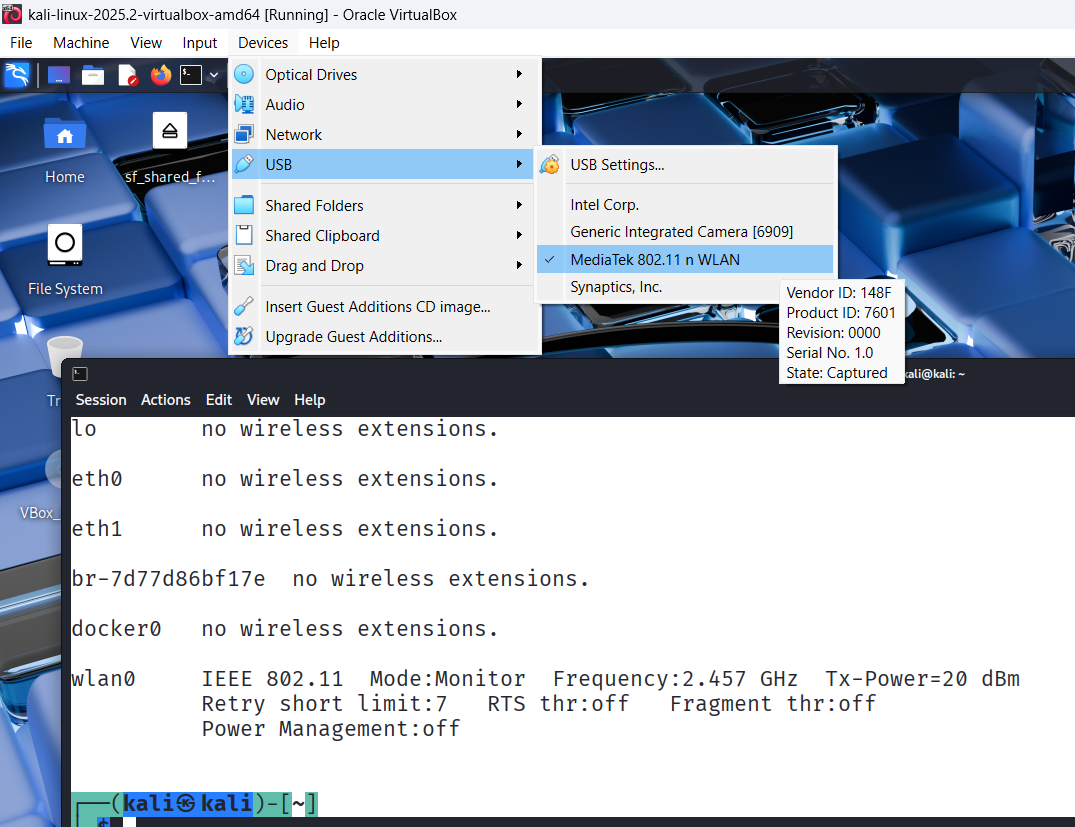}
	\caption{USB WiFi adapter used for monitor mode and packet injection (MediaTek MT7601U).}
	\label{fig:usbwirelessadapter}
\end{figure}

\begin{enumerate}
	
	\item \textbf{Initialisation and environment validation:}  
	The system is launched via a single CLI entry point (\texttt{cli.py}) with all execution parameters supplied explicitly by the operator, including the wireless interface identifier, scan duration, wordlist path, timeout, GenAI mode, and automation flags. Internally, argument parsing constructs a session configuration object that is propagated across all modules. A readiness phase validates kernel driver availability, USB device passthrough status, regulatory domain configuration, and the presence of required user-space tooling (e.g., \texttt{iw}, \texttt{airodump-ng}, \texttt{aireplay-ng}, \texttt{aircrack-ng}). The system also initialises a structured directory hierarchy for evidence storage and creates a timestamped session identifier to tag all subsequent artefacts. 
	This stage is illustrated in Fig.~\ref{fig:imgfirstsuccessrunpart1}, where the system banner and ``System Readiness Check'' are displayed immediately after invoking the wrapper script. 
	Internally, lightweight checks equivalent to querying interface capabilities (\texttt{iw dev}), tool availability (\texttt{which airodump-ng}), and USB enumeration are performed before execution continues.
	
	\item \textbf{Monitor-mode activation:}  
	After readiness confirmation, the system requests explicit user consent before enabling monitor mode on the selected interface. This transition is mandatory and intentionally enforced to guarantee deliberate engagement in RF operations. The interface state is programmatically re-queried after activation to confirm IEEE~802.11 monitor support, operating channel, transmit power, and MAC address. Network management services that could interfere with raw frame capture (e.g., \texttt{NetworkManager}, \texttt{wpa\_supplicant}) are temporarily suspended. 
	Fig.~\ref{fig:imgfirstsuccessrunpart2} shows the interface already operating in monitor mode, while Fig.~\ref{fig:usbwirelessadapter} confirms USB passthrough of the MediaTek MT7601U adapter inside VirtualBox. At the command level, this phase corresponds to actions conceptually equivalent to invoking monitor-mode utilities (e.g., \texttt{airmon-ng start}, or \texttt{ip link set} + \texttt{iw dev set type monitor}) followed by state verification via \texttt{iwconfig}.
	
	\item \textbf{Passive reconnaissance and client observation:}   
	The system then initiates a bounded passive scanning window, during which it listens for beacon frames and probe responses across channels to enumerate nearby APs. No frames are transmitted at this stage. For each discovered BSS (Basic Service Set), the system extracts ESSID/BSSID   pairs, channel number, encryption \&  authentication suites, RSSI, and observed client counts. Traffic rate and frame counters are monitored to approximate short-term activity levels and infer the probability of observing authentication handshakes in later phases. 
	This behaviour is shown in Fig.~\ref{fig:imgfirstsuccessrunpart2}, where ten networks are discovered and tabulated with encryption type, channel, power level, and associated stations. Internally, this corresponds to spawning a passive capture process similar to \texttt{airodump-ng} with channel hopping enabled, while continuously parsing its structured output stream.
	
	\item \textbf{Structured metadata aggregation:}  
	All raw observations collected during passive scanning are normalised into an internal schema capturing identifiers, protocol properties, signal metrics, and client activity indicators. The resulting dataset is serialised into structured JSON records and persisted to disk as part of the session evidence bundle. Raw capture files (PCAP/CAP) are retained locally for forensic inspection and reproducibility.

	\item \textbf{Prompt construction}
	\label{Step-GenAIPromptConstruction}
	
	The aggregated wireless metadata is serialised and injected into a predefined system prompt template that formalises the conditions under which GenAI reasoning is permitted. The template explicitly assigns the LLM the role of a ``seasoned wireless penetration tester'' in order to frame its reasoning style and domain assumptions, restricts the model to advisory analysis only (target ranking and feasibility assessment), and prohibits the generation of operational commands or autonomous actions. A deterministic JSON output schema is enforced, specifying fixed fields for candidate scores, justification, confidence estimates, and recommended targets, together with task-specific scoring criteria and interpretation guidelines to reduce ambiguity and improve consistency in dense wireless environments.
	
	To ensure traceability and experimental reproducibility, the prompt further embeds session context, including timestamps, session identifiers, tool name, and assessment stage. The authorised testing context and scope of engagement are also encoded directly within the prompt, together with legal, ethical, and operational constraints governing permissible behaviour.
	
	\item \textbf{Cost Estimation and human approval:}
	\label{Step-CostEstimation}
	Prior to  submission, the system computes the exact token count of the composed prompt and estimates the corresponding monetary cost for the selected model. These values are presented to the operator via an interactive ``LLM Budget Gate'', which pauses execution until explicit approval is provided (e.g., by responding \texttt{y}). Fig.~\ref{fig:imgfirstsuccessrunpart3} illustrates the cumulative token usage and cost table displayed at this stage.

	\item \textbf{LLM-based target ranking and feasibility assessment:}  
	Upon approval, the prompt is submitted to the GenAI backend. The model returns a single structured JSON object containing ranked candidate networks, confidence scores, inferred strengths and weaknesses, and a recommended strategy. The raw response is archived verbatim and then validated against a strict schema before being parsed into internal data structures. 
	Figs.~\ref{fig:imgfirstsuccessrunpart4} to~\ref{fig:imgfirstsuccessrunpart7} illustrate the returned JSON object, including ranked ESSIDs, BSSIDs, scores, and reasoning fields such as signal strength, encryption type, client count, and estimated handshake feasibility.
	
	\item \textbf{HITL target selection:}  
	The ranked candidate list is presented to the operator alongside the original passive scan table. The operator manually selects the target network by index. GenAI output is advisory only and never triggers wireless transmission or channel manipulation directly. 
	This interaction is visible in Fig.~\ref{fig:imgfirstsuccessrunpart6}, where the user selects the recommended candidate via numeric input. Internally, this simply updates the session context with the chosen BSSID and channel parameters.
	
	\item \textbf{Controlled active interaction and handshake capture:}  
	Only after explicit target selection does the system transition to controlled active operations. The wireless interface is locked to the target channel, a focused capture process is launched for the selected BSSID, and limited authentication-stimulating traffic may be generated to provoke protocol handshakes. All executed actions, timestamps, interface state changes, and process identifiers are logged. 
	Fig.~\ref{fig:imgfirstsuccessrunpart8} shows this phase, including channel locking, capture initialisation, and transmission attempts. At the tooling level, this corresponds to invoking focused capture utilities (e.g., \texttt{airodump-ng\footnote{Example invocation used in our experiments: \texttt{airodump-ng --bssid AA:BB:CC:DD:EE:FF --channel 6 --write target-01 wlan0}}}) and controlled injection utilities (e.g., \texttt{aireplay-ng\footnote{Example: \texttt{aireplay-ng --deauth 10 -a AA:BB:CC:DD:EE:FF wlan0}}}) while monitoring output streams for handshake indicators.
	
	\item \textbf{Handshake  validation and assessment:}   
	Captured authentication exchanges are analysed to verify structural correctness (e.g., message sequence completeness, replay counters, cipher suite consistency). Rather than recording a binary success flag, the system stores detailed validation metadata describing which protocol elements were observed and whether they satisfy formal handshake requirements.
	
	This step operates on the capture artefacts created in the previous phase (as shown in Fig.~\ref{fig:imgfirstsuccessrunpart8}) and produces structured validation records linked to the session identifier.

	When a target network is identified as operating under WPA or WPA2 using pre-shared keys (WPA/WPA2-PSK), the system explicitly verifies whether a complete and temporally consistent four-way handshake has been captured. In practice, this requires observing at least message~1 and message~2, and preferably all four messages, to ensure that the pairwise transient key (PTK) derivation can be validated offline. The capture file is inspected to confirm the presence of the \texttt{EAPOL} (Extensible Authentication Protocol over LAN) frames corresponding to the handshake sequence and to extract key parameters such as the AP-generated ANonce (Authenticator Nonce), the client-generated SNonce (Supplicant Nonce), MAC addresses, replay counters, and negotiated cipher suite. This validation step prevents false positives arising from partial captures, duplicated frames, or unrelated authentication attempts.
	
	If a valid handshake is confirmed, the system proceeds to assess passphrase resilience by invoking deterministic offline verification using standard tooling. The captured trace (e.g., \texttt{target-01.cap}) is supplied to a dictionary-based verification process using widely adopted tools such as \texttt{aircrack-ng}, \texttt{John~the~Ripper (JtR)}, or \texttt{hashcat}\footnote{Example command used in our experiments: \texttt{aircrack-ng -w /usr/share/wordlists/rockyou.txt target-01.cap}, which performs an offline dictionary-based verification attempt against the captured WPA/WPA2 four-way handshake using the standard \texttt{rockyou.txt} wordlist.}. 	
	 This tests potential candidate passphrases against the captured handshake without any further interaction with the target network. The outcome of this process (e.g., no match found, weak passphrase recovered, or verification inconclusive) is recorded as structured evidence together with the handshake metadata, timing information, and cryptographic parameters.

	\item \textbf{Evidence consolidation:}  
	All artefacts generated during the session are consolidated into a single evidence tree, including passive scan summaries, structured metadata JSON files, prompt and response transcripts, token usage logs, executed command traces, timestamps, and handshake validation reports. The directory structure preserves a strict one-session-per-folder layout to support later auditing and controlled replication.
	
	\item \textbf{GenAI-assisted report generation:}  
	From the consolidated evidence, a second prompt is constructed to generate a structured PenTesting report in JSON format. Sensitive fields (e.g., recovered secrets, raw packet payloads, client identifiers) are excluded or masked prior to submission. As in earlier stages, cost estimation and user approval are enforced before invoking the model.
	
	\item \textbf{Termination and archival:}  
	Finally, the session terminates either after successful completion, timeout expiration, or user interruption. The system restores network services if previously suspended and leaves all artefacts stored locally to support reproducibility, auditing, and later experimental analysis.
	
\end{enumerate}

\section{Discussion and Implications}
\label{DiscussionAndImplications}
We now discuss the benefits and limitations of \texttt{WiFiPenTester} and address the three research questions (RQs) posed in Section~\ref{Introduction}.

\subsection{Benefits and Features}
\label{BenefitsWiFiPenTester}

\texttt{WiFiPenTester} demonstrates that GenAI can be integrated into wireless PenTesting workflows in a controlled, auditable, and practically useful manner. By combining structured RF reconnaissance with model-assisted reasoning and strict human oversight, the system extends traditional wireless toolchains beyond static heuristics while preserving operational safety.

A central benefit of the system is its ability to perform GenAI-assisted target prioritisation under explicit human control. The integrated LLM analyses structured scan metadata and ranks candidate networks based on protocol configuration, signal quality, and observed activity. This significantly reduces the cognitive burden on operators when faced with dense wireless environments containing numerous APs and heterogeneous security configurations. Importantly, all recommendations remain strictly advisory and require explicit operator approval before any active operation is initiated.

The system further enforces strict HITL governance throughout the workflow. All disruptive or potentially intrusive actions — including monitor-mode activation, deauthentication, handshake capture, and report generation — are gated by explicit operator confirmation. This design choice reduces legal and ethical risk, prevents uncontrolled RF interference, and ensures that GenAI components cannot independently trigger wireless transmissions or active probing.

GenAI integration is additionally made budget-aware and fully auditable. Before each LLM interaction, the system computes token usage and estimates the associated monetary cost, presenting this information to the operator for approval. All prompts and responses are archived verbatim as part of the session evidence, enabling reproducibility, post-hoc analysis of model behaviour, and systematic experimentation with prompt design and reasoning strategies.

To support experimental rigour and operational accountability, the system employs structured evidence management. All artefacts — including passive scan outputs, normalised metadata, LLM transcripts, executed command traces, and protocol validation records — are consolidated into a session-scoped evidence tree. This organisation facilitates auditing, comparative evaluation across environments, and reproducible research workflows.

From a software engineering perspective, the system adopts a modular and extensible architecture. Implemented in Python with clearly separated components for wireless interaction, metadata processing, LLM communication, and reporting, the prototype can be extended to support additional wireless protocols, hardware adapters, LLM providers, or reporting formats with minimal redesign.

Finally, the prototype demonstrates operational realism by functioning on commodity hardware using standard wireless toolchains such as \texttt{aircrack-ng} and low-cost external USB adapters. This confirms the feasibility of the proposed approach under practical deployment constraints rather than relying on simulation, emulation, or synthetic traffic traces.

\subsection{Limitations and Risks}
\label{LimitationsRisksWiFiPenTester}

Despite its advantages, \texttt{WiFiPenTester} introduces several limitations and risks that must be acknowledged.

First, the system remains dependent on the quality and completeness of passive reconnaissance data. Wireless environments are inherently volatile: client mobility, fluctuating signal strength, hidden SSIDs, and transient associations can lead to incomplete or misleading snapshots. As a result, GenAI recommendations may be based on partial context, occasionally producing overconfident or suboptimal rankings.

Second, although strict human approval is enforced, active wireless operations such as deauthentication and handshake capture are intrinsically disruptive and may violate organisational policies or regulatory constraints if misused. The system mitigates this risk through governance controls, but it cannot eliminate the legal responsibility of the operator.

Third, the reliance on online LLM APIs introduces privacy and data-protection considerations. While raw packet contents and credentials are excluded from prompts, structured metadata may still reveal sensitive information about network topology or security posture. Deployment in production environments therefore requires careful compliance with institutional and legal data-handling policies. However, as discussed in Section~\ref{Privacy-Preserving GenAI Integration}, this limitation can be mitigated through strict prompt-level data minimisation and the use of locally deployed LLMs to eliminate external data transfer.

Fourth, the prototype has been evaluated primarily on WPA/WPA2 networks using dictionary-based verification. Full WPA3-SAE support remains limited in the current PoC and is left for future work.

Finally, the system inherits fundamental limitations of LLMs, including sensitivity to prompt phrasing, occasional hallucination, and inconsistent reasoning under distributional shift. Although structured prompting and schema validation reduce these risks, they do not eliminate them, reinforcing the necessity of continued HITL supervision.

\subsection{Answers to the Research Questions}
\label{AnswersToResearchQuestions}

We now address the three research questions posed in Section~\ref{Introduction}.

\begin{enumerate}
	\item \textbf{To what extent can GenAI assist in selecting viable wireless targets and attack strategies from reconnaissance data under human supervision?}
	
	Our results indicate that GenAI can provide meaningful and operationally useful support in prioritising wireless targets when supplied with structured scan metadata. The LLM consistently identified networks with stronger signal quality, active clients, and weaker authentication configurations as higher-feasibility candidates, often aligning with expert judgement. However, this assistance remains probabilistic rather than deterministic, and human validation is essential to account for environmental context that may not be captured in scan data (e.g., physical location, user intent, or legal scope). Thus, GenAI functions effectively as a decision-support layer rather than an autonomous attacker.
	
	\item \textbf{How does structured prompt engineering and decision framing influence the accuracy, consistency, and safety of LLM-driven recommendations in wireless PenTesting contexts?}
	
	Structured prompts with explicit output schemas, constrained task definitions, and protocol-aware framing substantially improved both the consistency and safety of model responses. In particular, requiring machine-readable JSON output and embedding operational boundaries reduced hallucinated actions and prevented the model from proposing unauthorised transmissions. Nevertheless, prompt sensitivity remained observable: small changes in metadata ordering or phrasing occasionally altered ranking priorities, highlighting the importance of systematic prompt design and validation in wireless contexts where incomplete information is common.
	
	\item \textbf{What practical limitations arise when deploying LLM-based reasoning within live, time-constrained, and non-deterministic RF environments?}
	
	Several limitations were observed. Temporal variability in client activity meant that networks deemed feasible during reconnaissance occasionally failed to yield handshakes during the active phase. Conversely, low-activity networks sometimes became viable unexpectedly. The LLM, operating on static snapshots, could not adapt to such dynamics without additional feedback loops. Furthermore, radio interference, channel congestion, and hardware driver behaviour introduced uncertainties that no purely symbolic reasoning system can fully anticipate. These findings suggest that GenAI integration is most effective when combined with continuous sensing and conservative operational thresholds rather than treated as a standalone decision engine.
	
\end{enumerate}

Overall, \texttt{WiFiPenTester} demonstrates that GenAI can augment wireless PenTesting in a principled and practically useful manner when embedded within a rigorously governed architecture. The system highlights both the promise of LLM-assisted reasoning for complex security workflows and the necessity of strict human oversight, protocol awareness, and evidence-driven validation in safety-critical cyber–physical environments.

\section{Conclusions and Further Research}
\label{ConclusionsAndFurtherResearch}

In this paper, we presented \texttt{WiFiPenTester}, a system for governed and reproducible GenAI-assisted wireless PenTesting. The system integrates large language model reasoning into the reconnaissance and decision-support phases of wireless security assessment, enabling intelligent target prioritisation, feasibility estimation, and strategy recommendation from structured scan metadata, while preserving strict human-in-the-loop   control over all active and potentially disruptive operations. We designed and implemented a proof-of-concept prototype operating on commodity hardware and standard wireless toolchains, and evaluated its behaviour across realistic wireless environments.

Our results demonstrate that GenAI can  reduce operator cognitive load during dense wireless reconnaissance, improve consistency in target selection, and provide structured, auditable reasoning to support tactical decision making. At the same time, the system enforces budget awareness, explicit user consent, protocol-aware validation, and comprehensive evidence logging, addressing key safety, legal, and reproducibility challenges inherent to wireless ethical hacking. Unlike prior automation tools, \texttt{WiFiPenTester} formalises decision support as a governed process rather than an opaque heuristic. 

Nevertheless, our findings also highlight important limitations. Model recommendations remain sensitive to prompt design and incomplete environmental context, wireless conditions introduce unavoidable non-determinism, and current support for WPA3-SAE is restricted to configuration and downgrade analysis rather than full attack automation. These constraints reinforce the necessity of continued human oversight and careful system governance when deploying GenAI within live radio-frequency environments.

Overall, \texttt{WiFiPenTester} represents a clear step forward towards safe, transparent, and operationally realistic GenAI-enabled wireless PenTesting. By embedding AI intelligence into the planning and evaluation stages — rather than execution itself — the system bridges the gap between manual expertise and automated assistance without compromising ethical or technical control.

Several promising directions for future research emerge from this work. First, full protocol-aware support for WPA3-SAE, including automated assessment of side-channel and implementation-level weaknesses, remains an open challenge and will be a primary focus of subsequent development. Second, extending the system to incorporate enterprise-scale features such as 802.1X/EAP analysis, rogue AP detection, and multi-AP correlation would significantly broaden its applicability in organisational environments. Third, adaptive prompt engineering strategies and online learning mechanisms could be explored to improve robustness under fluctuating wireless conditions and partial observations.

Further research will also examine quantitative human factors, including operator trust, decision accuracy, workload reduction, and time-to-target metrics when using GenAI-assisted workflows. Comparative benchmarking against existing wireless automation tools and emerging GenAI-based systems will help establish standardised evaluation baselines. Finally, ongoing work will investigate offline LLM deployment, cryptographic prompt integrity, and formal governance models to support secure adoption in regulated or sensitive operational contexts.

Taken together, these directions aim to advance \texttt{WiFiPenTester} from a proof-of-concept into a principled foundation for future GenAI-assisted wireless security assessment research.

\section*{Acknowledgements}
Some portions of this manuscript were refined using GenAI tools (specifically, ChatGPT) to assist with language polishing and structural clarity. Following the use of such GenAI tools, the authors thoroughly reviewed and edited the content as necessary and take full responsibility for the final publication. All core ideas, findings, experiments, and arguments were developed and authored by the named contributors.

\bibliographystyle{splncs04}
\bibliography{C:/Users/Dr_Ha/Desktop/Papers/database}



\begin{figure}[h]
	\centering
	\begin{tikzpicture}[
		node distance=0.9cm and 1.4cm,
		box/.style={
			rectangle, draw,
			minimum width=4.3cm,
			minimum height=0.8cm,
			align=center,
			font=\small
		},
		arrow/.style={-{Latex}, thick}
		]
		
		\node[box] (scan) {\textbf{WiFiPenTester}\\Passive Scan \& Metadata Aggregation};
		\node[box, below=of scan] (prompt) {\textbf{WiFiPenTester}\\Prompt Builder + Cost Estimator};
		\node[box, below=of prompt] (user1) {\textbf{User}\\Prompt Approval};
		\node[box, below=of user1] (llm1) {\textbf{LLM}\\Target Ranking \& Feasibility Reasoning};
		\node[box, below=of llm1] (parse) {\textbf{WiFiPenTester}\\Response Validation \& Parsing};
		\node[box, below=of parse] (user2) {\textbf{User}\\Target Selection Approval};
		\node[box, below=of user2] (attack) {\textbf{WiFiPenTester}\\Controlled Handshake Capture \& Assessment};
		
		\node[box, right=of llm1, xshift=0.5cm] (reportPrompt)
		{\textbf{WiFiPenTester}\\Report Prompt Builder\\(Facts Only / Masked Secrets)};
		
		\node[box, below=of reportPrompt] (user3)
		{\textbf{User}\\Report Prompt Approval};
		
		\node[box, below=of user3] (llm2)
		{\textbf{LLM}\\Structured PenTest Report\\Generation (JSON)};
		
		\node[box, below=of llm2] (log)
		{\textbf{Evidence Store}\\Logs, Prompts, Costs, Outcomes, Report};
		
		\draw[arrow] (scan) -- (prompt);
		\draw[arrow] (prompt) -- (user1);
		\draw[arrow] (user1) -- (llm1);
		\draw[arrow] (llm1) -- (parse);
		\draw[arrow] (parse) -- (user2);
		\draw[arrow] (user2) -- (attack);
		
		\draw[arrow] (attack.east) -- ++(0.6,0) |- (reportPrompt.west);
		
		\draw[arrow] (reportPrompt) -- (user3);
		\draw[arrow] (user3) -- (llm2);
		\draw[arrow] (llm2) -- (log);
		
	\end{tikzpicture}
	\caption{Operational architecture of \texttt{WiFiPenTester}}
	\label{WiFiTesterPPArchitecture2}
\end{figure}

\begin{figure}[h]
	\centering
	\includegraphics[width=1\linewidth]{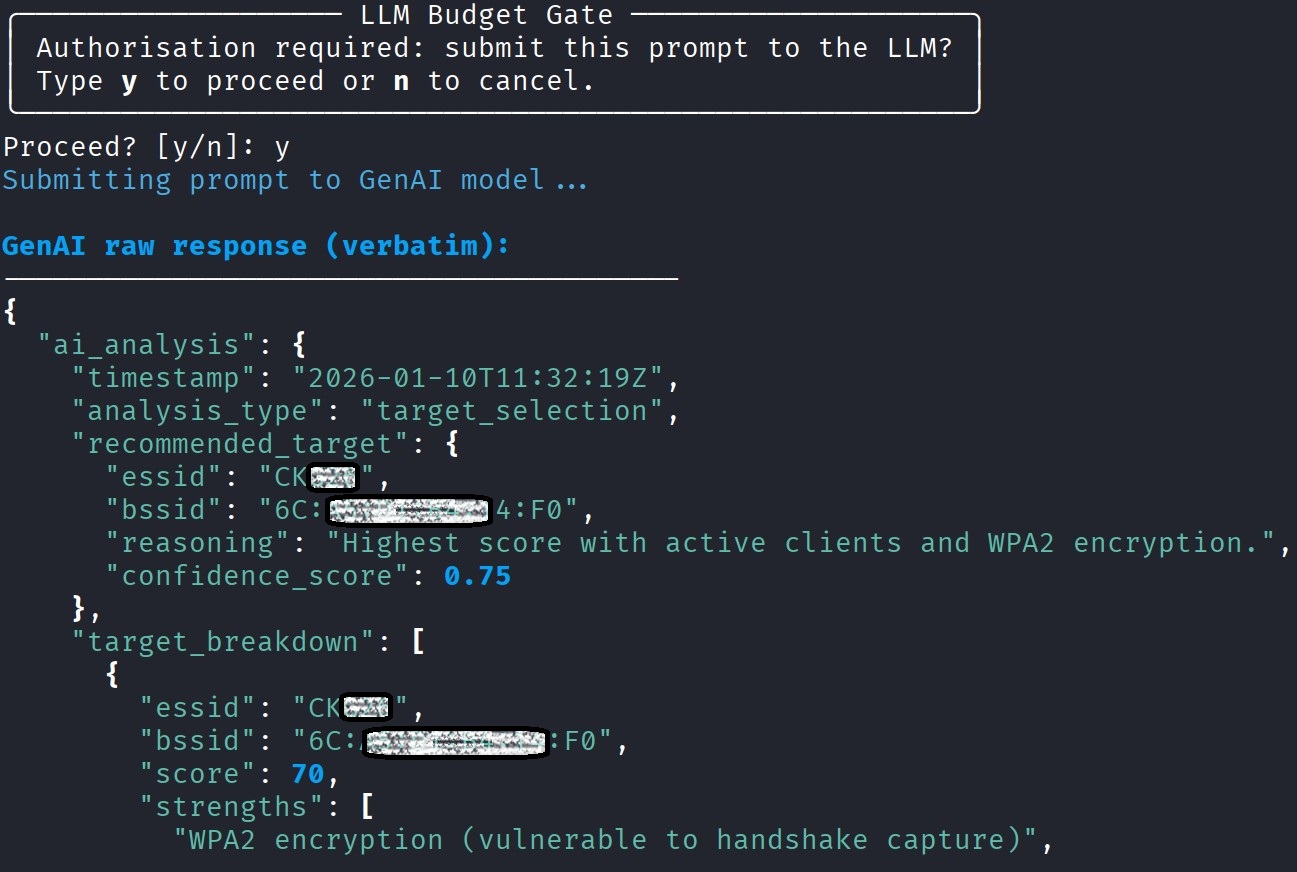}
	\caption{Structured target-selection prompt preview.}
	\label{fig:imgfirstsuccessrunpart4}
\end{figure}

\begin{figure}[h]
	\centering
	\includegraphics[width=1\linewidth]{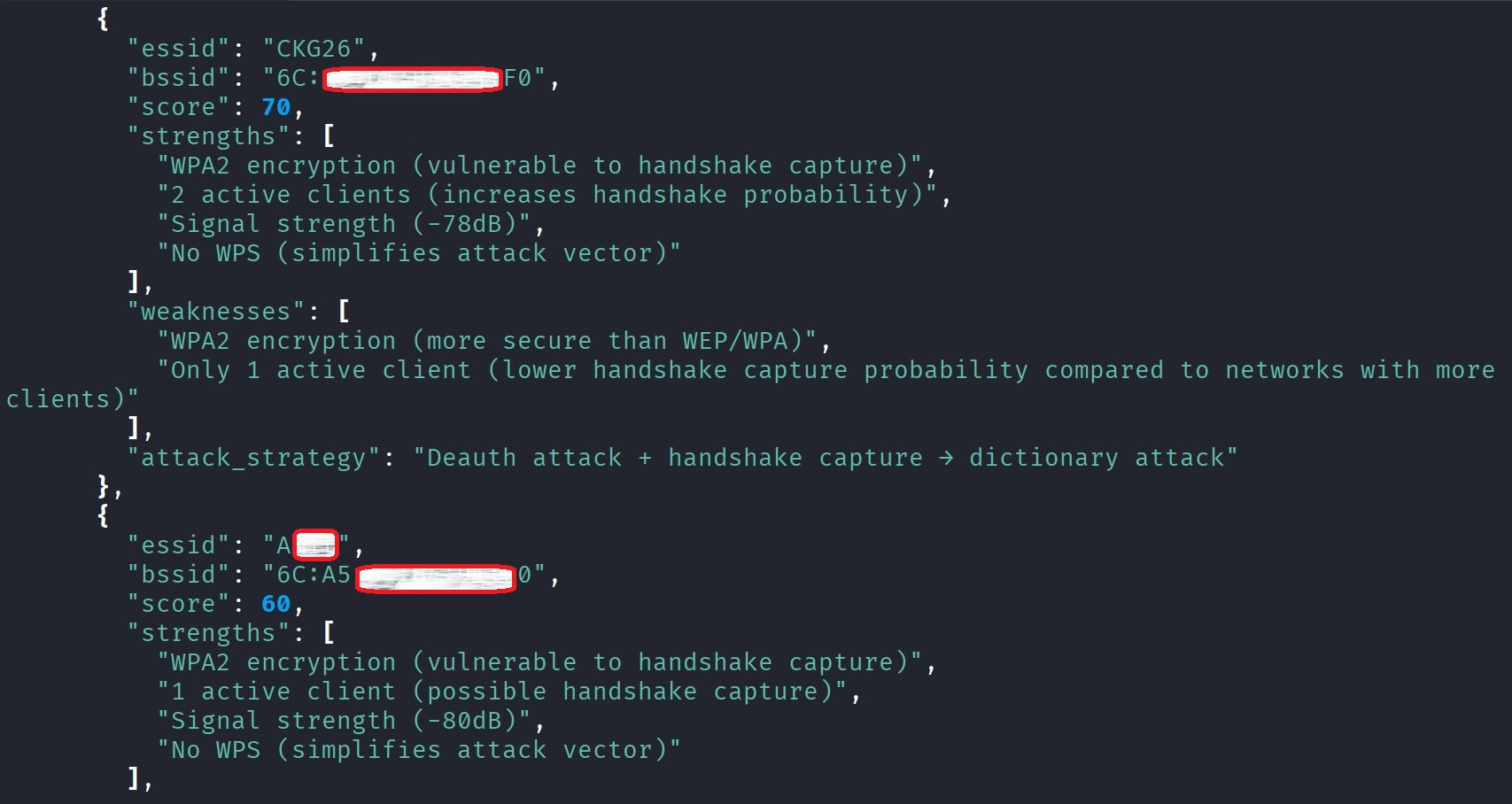}
	\caption{Structured target-selection prompt preview.}
	\label{fig:imgfirstsuccessrunpart5}
\end{figure}

\begin{figure}[h]
	\centering
	\includegraphics[width=1\linewidth]{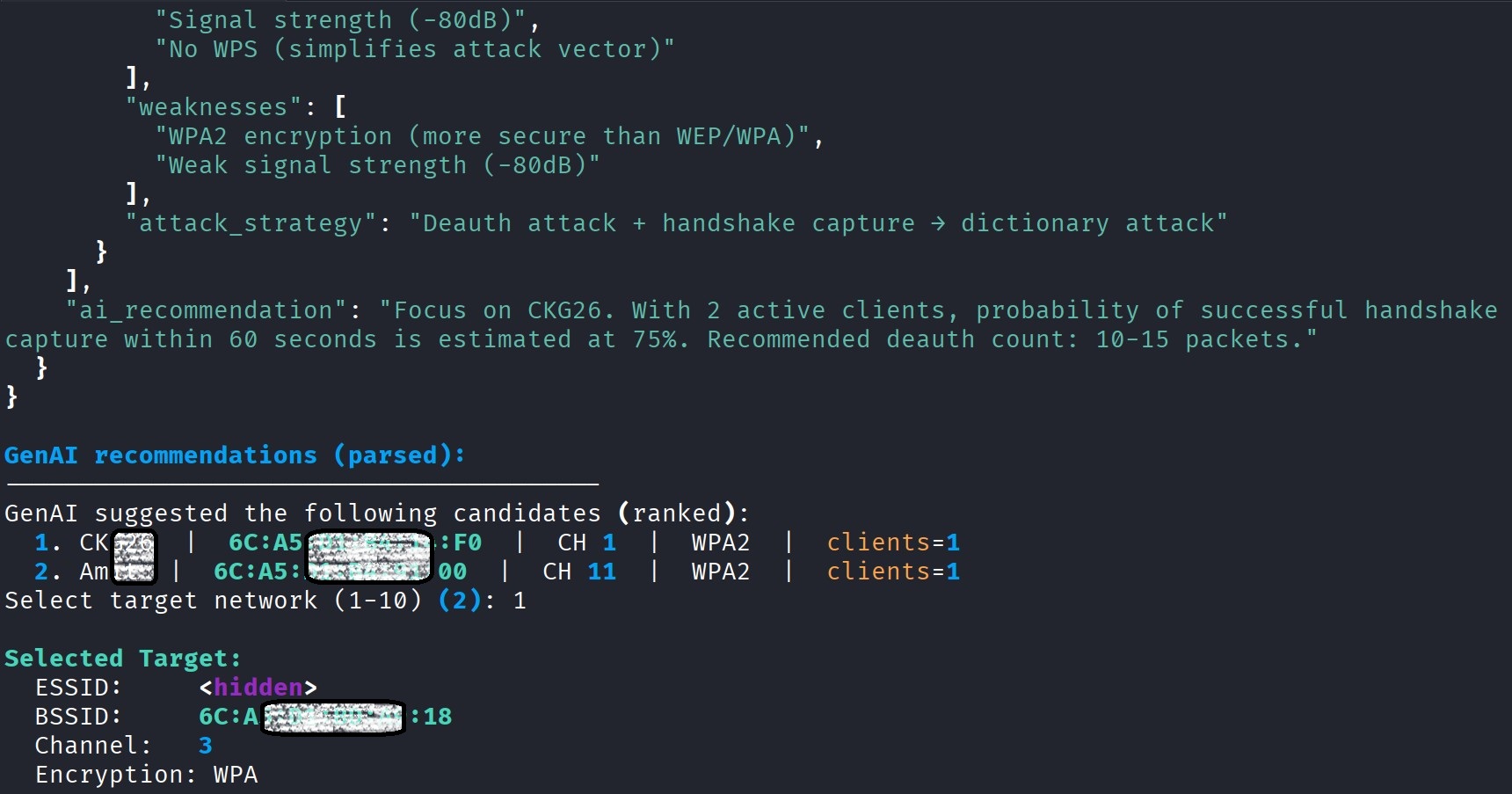}
	\caption{GenAI-based target ranking in structured JSON format.}
	\label{fig:imgfirstsuccessrunpart6}
\end{figure}

\begin{figure}[h]
	\centering
	\includegraphics[width=1\linewidth]{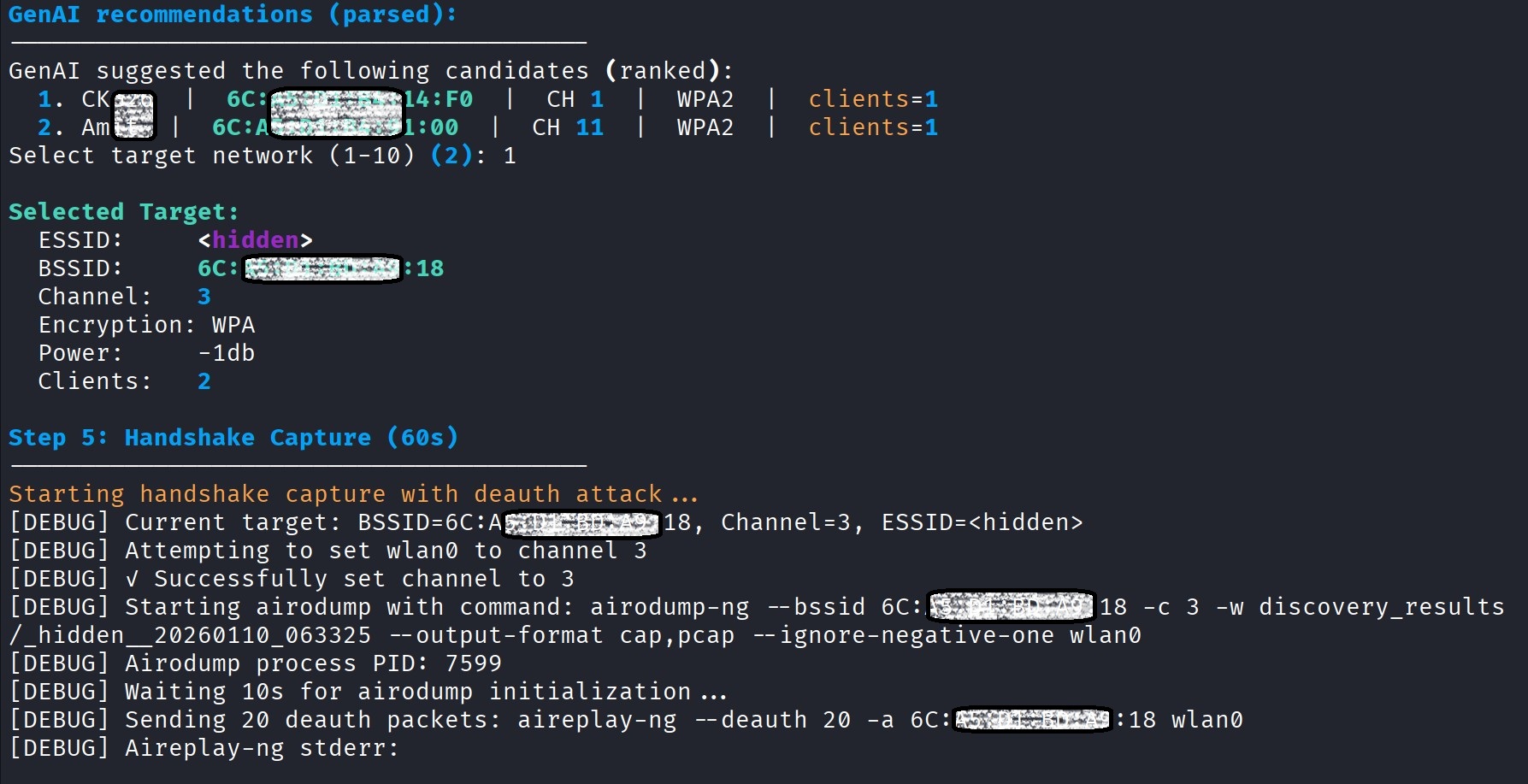}
	\caption{Human-in-the-loop target selection following GenAI assessment.}
	\label{fig:imgfirstsuccessrunpart7}
\end{figure}

\begin{figure}[h]
	\centering
	\includegraphics[width=1\linewidth]{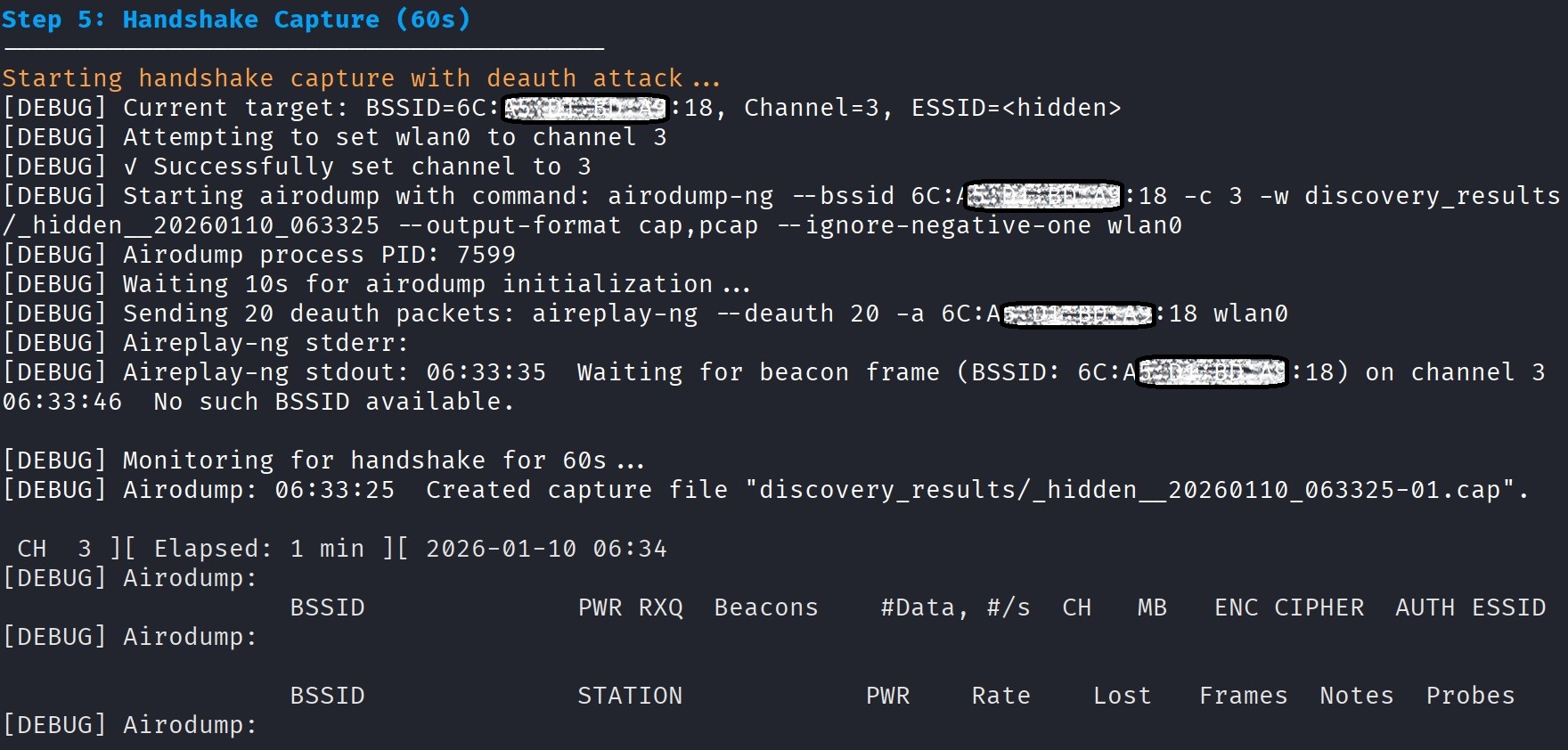}
	\caption{Controlled handshake capture for the selected target network.}
	\label{fig:imgfirstsuccessrunpart8}
\end{figure}

\end{document}